\documentclass[pdflatex,sn-basic]{sn-jnl}% Basic Springer Nature Reference Style

\usepackage{graphicx}
\usepackage[T1]{fontenc}
\usepackage[utf8]{inputenc}
\usepackage{amsmath}
\usepackage{amssymb}
\usepackage{booktabs}
\usepackage{array}
\usepackage{longtable}
\usepackage{multirow}
\usepackage{url}
\usepackage{fancyvrb}
\usepackage{listings}
\usepackage{xcolor}
\usepackage{geometry}
\usepackage{float} % Required for [H] float placement
\usepackage{hyperref}

\hypersetup{
    colorlinks=true,
    linkcolor=blue!70!black,
    citecolor=blue!70!black,
    urlcolor=blue!70!black,
    filecolor=blue!70!black,
    pdfborder={0 0 0},
    breaklinks=true
}

\setcitestyle{authoryear,open={(},close={)},citesep={;},aysep={,},yysep={,}}

\DefineVerbatimEnvironment{CodeChunk}{Verbatim}{
    fontsize=\footnotesize,
    frame=single,
    framesep=2mm,
    xleftmargin=3mm,
    xrightmargin=3mm,
    rulecolor=\color{gray!30}
}

\DefineVerbatimEnvironment{CodeInput}{Verbatim}{
    fontsize=\footnotesize,
    formatcom=\color{blue!70!black}
}

\DefineVerbatimEnvironment{CodeOutput}{Verbatim}{
    fontsize=\footnotesize,
    formatcom=\color{black!80}
}

\theoremstyle{thmstyleone}%
\theoremstyle{thmstyletwo}%

\theoremstyle{thmstylethree}%

\newcommand{\pkg}[1]{\texttt{#1}}
\newcommand{\code}[1]{\texttt{#1}}

\begin{document}

\title{Introducing '\texttt{inrep}': an R package that facilitates fully reproducible research workflows for survey-based assessments}

\author{
  \fnm{Clievins} \sur{Selva} 
  
  }

\affil{University of Hildesheim, Department of Psychology, Germany}


\abstract{
    Conducting research often involves managing multiple disconnected tools for survey design, data collection, response analysis, and report generation, leading to inefficiencies, increased error risks, and challenges in ensuring reproducibility. To address these issues, we introduce \texttt{inrep}, an open-source R package that integrates the entire assessment workflow within a unified, flexible framework in R. 
    With \texttt{inrep}, researchers can create customized assessments, streamline data management, and generate personalized participant reports without switching software or manually transferring data. \texttt{inrep} includes built-in support for generating structured prompts to guide large language models, enabling tailored adaptation of assessment components to specific study needs. 
    By consolidating all stages of the assessment process, \texttt{inrep} enhances research efficiency, improves reliability, and ensures full reproducibility, making sophisticated testing methodologies accessible to researchers, educators, and practitioners regardless of programming expertise.
}

\maketitle

%\section*{Introduction}\label{sec:introduction}

The landscape of empirical research is constantly evolving, with increasing demands for efficiency, transparency, and reproducibility in human data collection and analysis \citep{marwick2018, stodden2014a}. Yet, most researchers still contend with fragmented workflows: assessments are designed in one platform, data are collected in another, and analyses are performed elsewhere. This separation introduces inefficiencies, increases the risk of error, and limits transparency, which may contribute to persistent reproducibility challenges in psychology, education, and the social sciences \citep{baker2015, munafo2017, osc2015}. 

Most available tools support only isolated steps of the research workflow. Survey platforms focus on data collection but offer limited transparency and weak psychometric support. Statistical software handles analysis but does not assist with instrument design or data capture. Custom-built solutions bridge some of these gaps but require advanced skills, are hard to maintain, and are rarely reusable. These disjointed approaches lead to inefficiencies, raise the risk of errors, and exclude researchers without strong technical backgrounds \citep{marwick2018}. Additionally, there is a need for a unified, open-source solution to meet the growing demand for transparent, reproducible, and scalable research—that is, research workflows that can easily expand from small-scale pretests to larger studies, support repeated assessments (e.g., multiple waves), and adapt to frequent or longitudinal data collection as project needs grow.

One environment has predominantly emerged as the cornerstone of modern research: R\footnote{R is a free, open-source programming language and environment for statistical computing and graphics, widely used in research and data science. R packages are modular collections of functions, data, and documentation that extend R's capabilities for specific tasks or domains. A Shiny app is an interactive web application built using the \pkg{Shiny} R package, allowing users to interact with data and analyses through a graphical interface within the R environment. See \cite{R}.}. Numerous tailored R solutions exist, yet among these options, focus solely on the statistical analysis, while many capabilities of R remain underutilized by the broader research community. Critically, no package explicitly facilitates researchers to not only analyze data with R but also to record and present it within R.

\pkg{inrep} (short for ``instant reports'') directly addresses the gap of fragmented research workflows. The proposed software is an all-in-one R package that aims to streamline every stage of the assessment lifecycle. What sets \pkg{inrep} apart is its seamless integration within the R environment: Besides the by-design reproducible statistical analysis through R, researchers can design assessments, collect data, and generate participant-specific reports, all transparently, without leaving R or relying on proprietary services. While \pkg{inrep} supports both standard and adaptive assessment modes, this paper focuses on the standard workflow (adaptive features are described in the Supplementary Materials; see also \nameref{sec:notes}). 

Built on established psychometric foundations and designed in accordance with the FAIR for Research Software (FAIR4RS) principles \citep{lamprecht2020, katz2021}, \pkg{inrep} also leverages large language models\footnote{Large Language Models are systems trained to predict and generate sequences (text or code) by recognizing patterns in syntax, logic, and context from vast datasets.} (LLMs) to lower technical barriers. Users can translate plain-language research needs into analysis-ready workflows, fostering both reproducibility and accessibility. By consolidating all stages of the assessment process, \pkg{inrep} transforms research from a series of disconnected technical tasks into an integrated, transparent, and reproducible scientific workflow that sets a new standard for research in the age of artificial intelligence and open science.

\section*{Why \pkg{inrep} is Good Software}\label{sec:software-quality}

\begin{figure}[H]
    \centering
    \setlength{\fboxsep}{0pt}     % No padding between image and frame
    \setlength{\fboxrule}{0.8pt}  % Frame thickness
    \fbox{\includegraphics[width=1\textwidth]{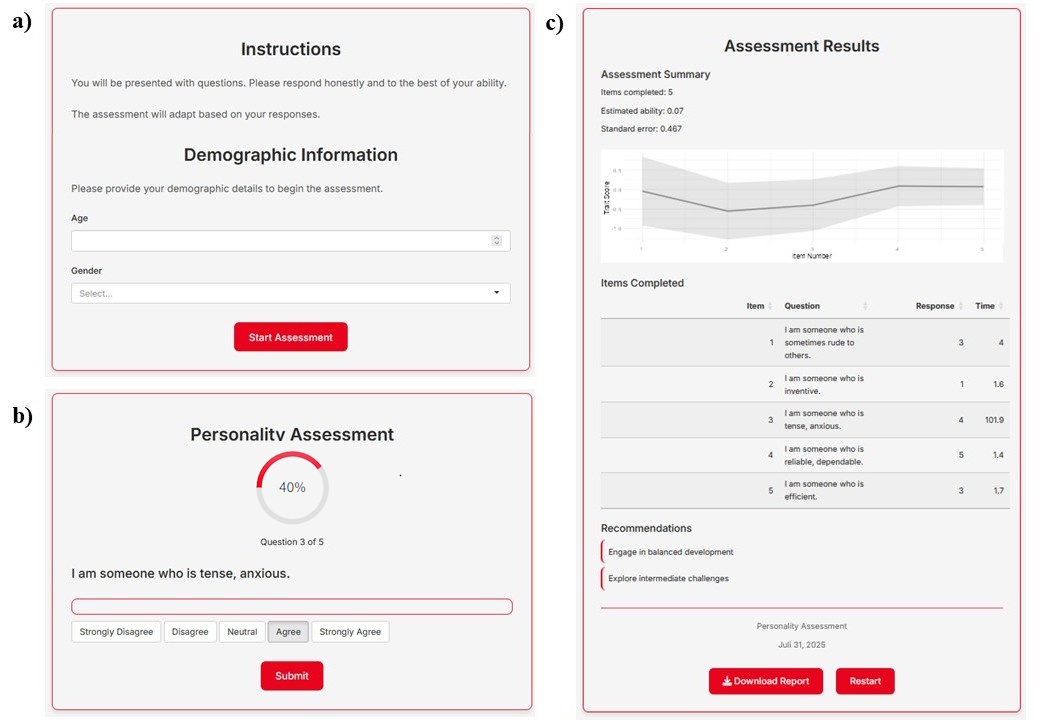}}

    \caption{\textbf{Participant-Facing Workflow in Adaptive Mode.} Panels show a minimal example of (\textbf{a}) onboarding with instructions and demographic variables, (\textbf{b}) the adaptive testing interface, and (\textbf{c}) the individualized participant feedback report. All components apply the \textit{Hildesheim} theme. Although displayed concurrently for illustrative purposes, participants experience these steps sequentially. See Listing~\ref{lst:adaptive-setup} in the Supplementary Materials for the implementation details. A cloud solution is hosted at \url{https://1q2zjr-selvastics.shinyapps.io/hilpub_inrep/}}
    
    \label{fig:inrep-report}
\end{figure}

Figure~\ref{fig:inrep-report} provides an overview of \texttt{inrep}'s participant-facing workflow, demonstrating how the package integrates onboarding, testing, and individualized feedback into a single, reproducible R environment. This unified approach is central to \texttt{inrep}'s design, addressing the fragmentation often encountered in empirical research workflows and making advanced psychometric methods and real-time reporting accessible to researchers across skill levels.

%\begin{figure}[H]
%    \centering
%    \includegraphics[width=1\textwidth]{dash.jpg}
%    \caption{\textbf{Participant-Facing Workflow in Adaptive Mode.} Panels show a minimal example of (\textbf{a}) onboarding with instructions and demographic variables, (\textbf{b}) the adaptive testing interface, and (\textbf{c}) the individualized participant feedback report. All components apply the \textit{Hildesheim} theme. Although displayed concurrently for illustrative purposes, participants experience these steps sequentially. See Listing~\ref{lst:adaptive-setup} in the Supplementary Materials for the implementation details. A cloud solution is hosted at \url{https://1q2zjr-selvastics.shinyapps.io/hilpub_inrep/}}
%    \label{fig:inrep-report}
%\end{figure}

%\footnotesize\textit{Note. Panels show a minimal example of (\textbf{a}) onboarding with instructions and demographic variables, (\textbf{b}) the adaptive testing interface, and (\textbf{c}) the individualized participant feedback report. All components apply the \textit{Hildesheim} theme. Although displayed concurrently for illustrative purposes, participants experience these steps sequentially. See Listing~\ref{lst:adaptive-setup} in the Supplementary Materials for the implementation details. A cloud solution of this is hosted at \url{https://1q2zjr-selvastics.shinyapps.io/hilpub_inrep/}}

Beyond this demonstration of integrated functionality, \pkg{inrep} is designed to meet the FAIR4RS principles, ensuring sustainability, discoverability, and collaboration \citep{lamprecht2020, katz2021}. Table \ref{tab:fair4rs} details how \pkg{inrep} implements these principles, enabling other researchers to likewise share, reuse, and build upon FAIR assessment workflows.

%In the context of the reproducibility crisis, where over 70\% of researchers have faced challenges in reproducing experiments \citep{baker2015}, \texttt{inrep} provides a solution by adhering to FAIR principles, which enhance the findability, accessibility, interoperability, and reusability of research software. 

\begin{table}[!p]
\thispagestyle{empty} % Hide page number on this page
\centering
\footnotesize
\caption{\pkg{inrep}'s implementation of FAIR4RS principles.}
\label{tab:fair4rs}
\begin{tabular}{p{0.35\linewidth} >{\arraybackslash}p{0.55\linewidth}}
\toprule
\textbf{FAIR Principle} & \textbf{Implementation in inrep} \\
\midrule
\textbf{Findable} & \\
F1: Software is assigned a globally unique and persistent identifier & The \pkg{inrep} package is hosted on GitHub (\href{https://github.com/selvastics/inrep}{https://github.com/selvastics/inrep}) with Zenodo archival for DOI assignment for its first major release (\href{https://doi.org/10.5281/zenodo.16682020}{https://doi.org/10.5281/zenodo.16682020}), ensuring persistent global identification. \\
F1.1: Components of the software representing levels of granularity are assigned distinct identifiers & Core components have distinct identifiers: the main package (\pkg{inrep}), example datasets (\code{bfi\_items}, \code{cognitive\_items}, \code{math\_items}), and modular functions (\code{launch\_study}, \code{create\_study\_config}, \code{estimate\_ability}) are individually documented and versioned. \\
F1.2: Different versions of the software are assigned distinct identifiers & Each release follows semantic versioning (currently v1.0.0) with Git tags and Zenodo DOIs for major releases, ensuring version-specific identification. \\
F2: Software is described with rich metadata & Comprehensive \texttt{DESCRIPTION} file includes detailed metadata (title, authors, version, license, dependencies, system requirements), supplemented by extensive \texttt{README}, vignettes, and function documentation via \pkg{roxygen2}. \\
F3: Metadata clearly and explicitly include the identifier of the software they describe & All metadata files (\texttt{DESCRIPTION}, \texttt{CITATION}, \texttt{README}) explicitly reference the package name \pkg{inrep} and GitHub repository URL, with DOI integration for Zenodo releases. \\
F4: Metadata are FAIR, searchable and indexable & Machine-readable metadata in standardized R package format, indexed on GitHub, with planned \pkg{CRAN} submission for broader discoverability. \\
\midrule
\textbf{Accessible} & \\
A1: Software is retrievable by its identifier using a standardised communications protocol & Retrievable via HTTPS through GitHub using standard R installation commands (\code{devtools::install\_github("selvastics/inrep")}), with planned \pkg{CRAN} availability for \code{install.packages()} access. \\
A1.1: The protocol is open, free, and universally implementable & Uses open protocols: HTTPS for web access and Git for version control, both free and universally accessible without proprietary requirements. \\
A1.2: The protocol allows for an authentication and authorization procedure, where necessary & Public GitHub repository requires no authentication; standard OAuth2 authentication available for contributors. Future private institutional deployments can leverage existing Git authentication mechanisms. \\
A2: Metadata are accessible, even when the software is no longer available & Repository metadata persisted through GitHub's infrastructure and Zenodo archival ensures long-term metadata accessibility independent of active maintenance status. \\
\midrule
\textbf{Interoperable} & \\
I1: Software reads, writes and exchanges data in a way that meets domain-relevant community standards & Utilizes standard R data structures (data.frames, lists) and supports industry-standard formats (CSV, JSON, RDS). Follows R community conventions. \\
I2: Software includes qualified references to other objects & All dependencies explicitly specified in \texttt{DESCRIPTION} with version constraints. References external psychometric literature, datasets with proper citations. Further, \pkg{inrep} is essentially a wrapper that integrates with the \pkg{TAM} package \cite{robitzsch2024} for psychometric computations and the \pkg{Shiny} package \cite{shiny} for app applications. \\
\midrule
\textbf{Reusable} & \\
R1: Software is described with a plurality of accurate and relevant attributes & Rich documentation ecosystem: detailed \texttt{DESCRIPTION} metadata, comprehensive \texttt{README} with usage examples, multiple vignettes covering workflows, \pkg{roxygen2}-documented functions, and academic preprint describing methodology and applications. \\
R1.1: Software is given a clear and accessible license & Released under \texttt{MIT License} clearly stated in \texttt{LICENSE} file and \texttt{DESCRIPTION}, providing maximum reusability while maintaining attribution requirements. \\
R1.2: Software is associated with detailed provenance & Complete Git version control history documenting all changes, author contributions tracked through commits, and release documentation on Zenodo with contributor metadata and timestamps. \\
R2: Software includes qualified references to other software & Dependencies listed in \texttt{DESCRIPTION} with specific version ranges; documentation includes proper citations to \pkg{TAM} package, \pkg{Shiny} framework, and relevant psychometric literature with full bibliographic references. \\
R3: Software meets domain-relevant community standards & Adheres to R package development standards (\pkg{roxygen2} documentation, \pkg{testthat} framework, R CMD check compliance), follows psychometric software conventions, and implements accessibility guidelines (i.e., compliance with Web Content Accessibility Guidelines) for educational assessment tools. \\
\bottomrule
\end{tabular}
\vspace{0.5em}
\begin{flushleft}
\footnotesize\textit{Note.} FAIR4RS refers to the Findable, Accessible, Interoperable, and Reusable principles for research software. See \cite{lamprecht2020} and \cite{katz2021} for more details. 
\end{flushleft}
\end{table}

% Moreover, two features illustrate why \pkg{inrep} stands out as good research software. First, it enables researchers to conduct all steps of their study, from design to reporting, within a single, free, and open-source environment. This not only streamlines the workflow but also ensures that every step is fully reproducible and transparent. Second, \pkg{inrep} automatically generates individualized result reports at the end of each session, tailored to study objectives and customizable in appearance (e.g., themes for colorblind or dyslexic participants). This supports open-science principles, enhances participant engagement, and enables feedback-driven research and education.
\subsection*{How \texttt{inrep} Supports Fair and Equitable Assessment}\label{sec:fair-assessment}

Equitable research requires thoughtful attention to participant experience and accessibility. Traditional assessment platforms often overlook the diverse needs of participants, particularly those from marginalized communities or with different learning preferences. \texttt{inrep} addresses these challenges through comprehensive design features that prioritize participant comfort, engagement, and representation.
\pkg{inrep} includes multiple built-in themes designed to support diverse participant needs. The Inrep theme provides a clean, professional interface, while specialized themes accommodate participants with visual impairments, dyslexia, or color vision differences. These themes adjust not only colors and contrast but also font choices, spacing, and layout to ensure accessibility across different populations. Researchers can easily select appropriate themes based on their participant demographics or allow participants to choose their preferred interface.

Beyond visual accessibility, \texttt{inrep} supports multiple languages and cultural adaptations, enabling researchers to conduct studies across diverse populations without technical barriers. The platform also includes features for clear instructions, progress indicators, and user-friendly navigation that reduce cognitive load and participant fatigue. These features are particularly important when working with vulnerable or stressed populations.

Quality control features are seamlessly integrated into the workflow, helping researchers maintain ethical standards without compromising participant experience. Automated data validation and real-time monitoring ensure data integrity while respecting participant time and effort. For researchers interested in adaptive features that can further optimize participant experience and reduce assessment burden, these capabilities are particularly important when working with vulnerable or stressed populations.

These accessibility and equity features make \texttt{inrep} a powerful tool for inclusive research across disciplines, supporting the principles of open science and social responsibility while ensuring that research benefits all participants \citep{munafo2017}.

\subsection*{What LLMs can do for \pkg{inrep}}

Because studies vary widely in their research questions, methodologies, target populations, and analytical goals, a one-size-fits-all software solution is impractical. The \texttt{inrep} package addresses this complexity by leveraging LLMs to provide researchers with adaptable, customizable workflows. This integration facilitates personalized research approaches without requiring advanced programming skills.

\pkg{inrep} enables context-sensitive study design with functions such as \code{llm\_prompt} and \code{llm\_assistance}. These functions generate structured prompts for external LLMs (e.g., ChatGPT, Perplexity) that help users adapt assessment components to their specific study needs. Users can receive LLM-generated guidance or code suggestions that can be easily integrated into their \pkg{inrep} workflows.

This streamlined approach allows researchers to express their requirements in plain language and receive tailored support for implementing their studies. The integration empowers both newcomers and experienced users to efficiently design and deploy assessments while maintaining full control over their research process.

However, \pkg{inrep} ensures responsible LLM integration through structured templates, validation mechanisms, and transparent human-in-the-loop workflows. This approach ensures that LLMs augments rather than replaces researcher judgment, maintaining the highest standards of rigor, reproducibility, and ethical responsibility.

\section*{Discussion}\label{sec:disc}

This paper introduces \pkg{inrep}, an open-source R package designed to unify and streamline workflows for survey-based research. By integrating study design, data collection, analysis, and reporting, \pkg{inrep} directly addresses pervasive replicability challenges highlighted by Baker in her survey of 1,576 scientists \citep{baker2016}.

Baker's findings reveal a critical replicability crisis: over 70\% of researchers failed to replicate another scientist's work, and more than half could not replicate their own experiments \citep{baker2016}. These results, drawn from diverse scientific disciplines, suggest that fragmented workflows and insufficient transparency throughout the research process—including data analysis—may be fundamental obstacles to reliable research \citep{marwick2018}. However, the concept of replicability itself remains debated, with questions about what aspects of research must be replicable and whether deviations in design and measurement methods that still yield similar results might actually strengthen theories \citep{gignac2016}.

The replication crisis in psychology and related disciplines stems from interconnected methodological, cultural, and systemic factors \citep{baker2016, munafo2017, scheel2021}. While no single solution can fully resolve this complex challenge, \pkg{inrep} represents a significant step toward more replicable research by directly addressing workflow fragmentation. By unifying assessment design, data collection, analysis, and reporting within a transparent, FAIR-compliant framework, \pkg{inrep} mitigates critical vulnerabilities such as undocumented analytical decisions, manual data transfer errors, and inaccessible proprietary tools. However, achieving truly replicable science requires complementary efforts across the research ecosystem, including stronger theoretical foundations, improved statistical training, and institutional incentives for open practices. \pkg{inrep} thus serves as both a practical tool for researchers and a catalyst for broader methodological reform, demonstrating how integrated technical solutions can meaningfully advance replicability while acknowledging that comprehensive solutions must address the crisis's multifaceted nature.

As noted in recent critiques of psychological science, many replication failures may stem not from methodological flaws alone, but from insufficiently developed theories that fail to specify precise mechanisms or boundary conditions \citep{meehl1990, yarkoni2020}. While \pkg{inrep} cannot directly address these theoretical challenges, it mitigates methodological vulnerabilities by embedding robust, transparent practices into routine research workflows. The package lowers technical barriers while promoting methodological rigor \citep{stodden2014a, stodden2014b, lamprecht2020}. Its unified open-source framework consolidates all research stages, directly countering pitfalls like poor documentation and disconnected processes identified in replicability studies \citep{baker2015, baker2016}. By fostering accountability and continuous improvement, \pkg{inrep} responds to growing demands for transparency in science \citep{lamprecht2020, katz2021}. We view such advancements as essential for enhancing research integrity across disciplines.

\subsection*{Core Contributions of \pkg{inrep}}

The workflow illustrated with \pkg{inrep} introduces three critical advances that collectively advance empirical research practice. 

\textbf{It is Reproducible.} By consolidating all research stages within R, \pkg{inrep} eliminates the inefficiencies and error risks associated with manual data transfers between platforms. Every aspect of the research process---from initial survey configuration through final reporting---is automatically documented, creating comprehensive audit trails that enable methodological verification and replication. This addresses reproducibility concerns by making transparent methodology the default rather than an exceptional effort \citep{stodden2014a, stodden2014b}.

\textbf{It is Customizable.} \pkg{inrep} is designed to support inclusive research through flexible accessibility features. Users can adapt the interface with themes tailored to visual impairments or dyslexia, enable multilingual options, and apply culturally responsive modifications. Instant participant reporting is also fully adjustable, allowing researchers to tailor feedback to diverse audiences. These customizations promote ethical and socially responsible practices by ensuring that study designs can be aligned with participant needs.

\textbf{It is Free.} Commercial survey tools provide robust data collection capabilities, but they are often expensive, offer only limited analytical functionality, and lack algorithmic transparency. The \pkg{inrep} package is completely free and open-source, making it accessible to researchers across disciplines and institutional contexts. This democratizes access to advanced psychometric methods and real-time reporting, enabling researchers without extensive programming expertise to conduct high-quality, reproducible research.

\subsection*{Responding to Evolving Educational Challenges with \pkg{inrep}}

The rapid integration of artificial intelligence (AI) into empirical research is fundamentally reshaping methodological standards in education and the social sciences. As AI-driven tools become increasingly prevalent, researchers face new challenges in maintaining core scientific values such as validity, reliability, and representativeness. These developments demand a careful re-examination of established practices to ensure that innovation does not come at the expense of transparency or rigor.

The \texttt{inrep} framework is designed as a direct response to these evolving challenges. By embedding adaptive testing and automated fairness diagnostics within the familiar R analytical environment, \texttt{inrep} breaks down traditional barriers between data collection and analysis. This integration empowers researchers to uphold rigorous psychometric standards throughout every stage of their workflow, even as assessments are dynamically tailored to individual respondents. Importantly, \texttt{inrep} achieves this without sacrificing transparency or replicability, directly addressing concerns about the opacity and unpredictability of AI-driven methods.

Through its unified approach, \texttt{inrep} enables researchers to monitor, validate, and document every step of the assessment process. This not only preserves the validity and precision of measurement, but also ensures that adaptive and AI-supported methods remain interpretable and open to scrutiny. In doing so, \texttt{inrep} supports the responsible adoption of AI in empirical research, providing a robust infrastructure for methodological innovation that remains grounded in the foundational principles of scientific inquiry.

A critical design philosophy underlying \texttt{inrep} is the responsible integration of AI with well-established empirical practices. Rather than treating LLMs as black-box solutions where data are indiscriminately funneled for processing, a practice that risks violating methodological standards, \texttt{inrep} situates AI as a complement to, not a replacement for, traditional workflows. This hybrid approach maintains foundational criteria such as validity, reliability, and interpretability, ensuring that AI’s efficiencies are harnessed without eroding the quality or integrity of research. Careful model selection, prompt engineering, and integrated opportunities for empirical validation enable researchers to monitor and scrutinize AI outputs within robust psychometric frameworks.

Furthermore, \texttt{inrep} exemplifies how LLMs can enhance human expertise by integrating LLMs as controlled, transparent assistants within the established, reproducible framework of R. Rather than replacing researcher judgment, \texttt{inrep} leverages LLMs to streamline workflows through guided code generation and methodological support, always leaving ultimate oversight with the user. This human-in-the-loop approach preserves rigor, reproducibility, and ethical accountability by grounding AI contributions in validated methods and peer-reviewed knowledge. As the field of educational research embraces AI, \texttt{inrep} exemplifies responsible use of AI tools with the integrity and transparency that are essential for socially responsive and methodologically robust scholarship.

Finally, \texttt{inrep} transcends methodological dichotomies by facilitating continuous data quality monitoring and adaptive, inclusive testing procedures, particularly for underrepresented or marginalized populations. Such functionality not only broadens the scope of what can be achieved automatically but also advances the imperative for socially responsible research design. Features such as comprehensive provenance tracking and outputs compliant with FAIR principles reinforce ethical accountability and provide a clear audit trail, setting a new standard for reproducibility in measurement.

\subsection*{Limitations and Future Work}

The transition to unified research workflows does present certain challenges that warrant acknowledgment. Researchers unfamiliar with R may initially encounter a learning curve, though the package's AI-assisted guidance specifically addresses this concern by enabling plain-language interaction with complex methodological workflows. The initial setup investment required for fully reproducible research infrastructure is also higher than traditional point-and-click alternatives, but this reflects the fundamental requirements of building sustainable, transparent research processes rather than a design limitation.

Technical constraints around very large-scale deployments are being addressed through ongoing optimization efforts and infrastructure improvements. These limitations are balanced against the significant advantages of unified workflows, complete methodological transparency, and adherence to FAIR4RS principles \citep{lamprecht2020, katz2021}.

Future development priorities focus on expanding accessibility features, enhancing scalability, and integrating emerging methodological innovations while maintaining the package's core commitments to transparency, reproducibility, and ethical accountability.

\subsection*{Conclusion}

In conclusion, \pkg{inrep} offers a practical solution to longstanding challenges in empirical research while advancing broader goals of open science and research equity. By unifying workflows, democratizing methodologies, and prioritizing transparency, the package provides a roadmap for conducting rigorous, reproducible, and socially responsible research in an increasingly complex scientific landscape.

\subsection*{Conflicts of Interest}
The author declares no conflicts of interest related to this work.

\subsection*{Correspondence}

Correspondence concerning this article should be addressed to Clievins Selva, University of Hildesheim, Department of Psychology, Universitätsplatz 1, 31141 Hildesheim, Germany. Email: \href{mailto:clievins.selva@uni-hildesheim.de}{clievins.selva@uni-hildesheim.de}

\subsection*{Funding}

No external funding was received for this work.

\subsection*{Notes}
\label{sec:notes}

\begin{itemize}
  \item \textbf{\pkg{inrep} is still in development.} If you are using \pkg{inrep} in a version below 2.00, please contact the author for package support, as the package is still under active development.
  \item \textbf{Versioning and reporting:} Always report the \pkg{inrep} version and dependencies used in your study. In adaptive mode, please always cite \pkg{TAM} for the psychometric computations.
  \item \textbf{Data privacy and security:} \pkg{inrep} supports both local and remote (Shiny Server) deployment. Users are responsible for ensuring compliance with data protection regulations when handling participant data, especially in cloud or institutional settings.
  \item \textbf{LLM integration:} \pkg{inrep} provides functions for LLM-assisted workflows, but users must review and validate any AI-generated code or analysis. LLM outputs should not be used blindly for critical research decisions.
  \item \textbf{Accessibility:} \pkg{inrep} aims for accessibility, but users deploying custom interfaces should test for accessibility in their specific context.
  \item \textbf{Community contributions:} The \pkg{inrep} package is open-source and welcomes contributions from the community. Users are encouraged to share their enhancements, bug fixes, and feature requests via the GitHub repository.
    \item \textbf{Existing tools for adaptive testing and psychometrics:} The following R/Shiny tools are available for related purposes, but differ from \pkg{inrep} in important ways:
        \begin{itemize}
          \item \textbf{mirtCAT}: A Shiny application for computerized adaptive testing (CAT) built on the \pkg{mirt} package \citep{chalmers2012mirt}, which implements multidimensional item response theory models. \pkg{mirtCAT} \citep{chalmers2016mirtcat} is a well-established tool for CAT in R, allowing interactive adaptive testing based on the \pkg{mirt} framework. However, \pkg{mirt} does not provide functionality for plausible value estimation. In contrast, \pkg{inrep} integrates with the \pkg{TAM} package \citep{robitzsch2024}, an R package allowing for plausible value estimation and latent variable modeling in large-scale educational and psychological research. This integration allows \pkg{inrep} to offer adaptive testing features alongside robust plausible value methods, supporting more comprehensive psychometric analyses. Unlike \pkg{mirtCAT}, \pkg{inrep} unifies adaptive assessment, data collection, psychometric modeling, and reporting within a single reproducible R workflow.
          \item \textbf{ShinyTAM}: An interactive web app for Rasch-model psychometric analysis using a Shiny interface. It does not support adaptive testing and is focused on classical Rasch analysis \citep{shinytam}.
          \item \textbf{catIRT Tools}: A Shiny application for adaptive testing and IRT simulations, based on the \pkg{mirt} and \pkg{catR} packages (not TAM). It is designed for IRT modeling and simulation, not for full CAT workflow management \citep{ogura2020catirt}.
          \item \textbf{RSCAT}: An R package and Shiny app for computerized adaptive testing using the shadow-test approach, built on 3PL models (not TAM). It is intended for a specific type of adaptive testing and does not provide the same workflow as \pkg{inrep} \citep{rscat}.
        \end{itemize}
\end{itemize}

\subsection*{Supplementary Materials}

\subsubsection*{Software Availability and Installation}

\pkg{inrep} is an open-source R package designed to support the development, deployment, and analysis of reproducible assessment workflows entirely within the R environment. The full source code, documentation, and supplementary materials are publicly available on GitHub at:  
\url{https://github.com/selvastics/inrep}

To install the latest version of the package, use the \pkg{devtools} package in R:

\newpage

\begin{lstlisting}[language=R, caption={Installation}]
# Install from GitHub
devtools::install_github("selvastics/inrep") 

# Load the package
library(inrep)
\end{lstlisting}

\subsubsection*{Basic Assessment Setup}

A simple fixed-form assessment can be launched with just a few lines of code:

\begin{lstlisting}[language=R, caption={Basic assessment setup}]
library(inrep) 
data(bfi_items) # Built-in 30-item personality inventory

# Configure the study
config <- create_study_config(
  name = "Personality Assessment",      # Study name
  items = bfi_items,                    # Item pool
  demographics = c("Age", "Gender"),    # Demographic fields
  theme = "hildesheim"                  # UI theme
)

# Launch the assessment
launch_study(config)
\end{lstlisting}

\subsubsection*{Adaptive Assessment Features}\label{sec:adaptive-supplement}

For researchers interested in advanced capabilities, \texttt{inrep} supports sophisticated adaptive testing through integration with established psychometric packages.

In adaptive mode, \texttt{inrep} leverages item response theory (IRT) to implement adaptive testing, dynamically selecting items based on a participant's estimated ability (or any other objective) in real time \citep{embretson2000}. This approach can reduce test length compared to fixed-form assessments, minimizing participant burden and fatigue, particularly beneficial for diverse populations \citep{wainer2000}.

%\textbf{Technical Implementation Details:} In adaptive mode, \texttt{inrep} can apply different objectives for item selection like theta or standard error thresholds—for instance, one could implement a median split on ability estimates—to assign tailored item sets dynamically. Participants below a threshold receive simpler items, while those above receive more challenging ones, optimizing measurement precision and participant experience. The system also supports dynamic group switching, allowing participants who demonstrate higher ability during simpler assessments to transition to advanced item sets.

\begin{lstlisting}[language=R, caption={Adaptive assessment setup}, label={lst:adaptive-setup}]
library(inrep) 
data(bfi_items) # Built-in 30-item personality inventory (with item parameters)

# Configure adaptive study
config <- create_study_config(
  name = "Adaptive Personality Assessment",
  model = "GRM",                        # Graded Response Model for Likert items
  max_items = 15,                       # Maximum items to administer
  min_items = 5,                        # Minimum items before stopping
  min_SEM = 0.3,                        # Minimum standard error for stopping
  demographics = c("Age", "Gender"),
  theme = "hildesheim"
)

# Launch adaptive assessment
launch_study(config, bfi_items)
\end{lstlisting}

%\bibliography{references}
{\footnotesize
\bibliography{references}
}

\end{document}